\documentclass[aps,12pt,msam,msbm]{iopart}

\usepackage{graphicx}
\usepackage{amssymb}
\usepackage{wasysym}
\usepackage{verbatim}
\usepackage{color}
\def\newblock{}
\newcommand{\Tbkt}{{T_\mathrm{BKT}}}
\newcommand{\nn}[1]{{\langle{#1}\rangle}}
\newcommand{\hc}{\mathrm{h.c.}}
\newcommand{\Eqref}[1]{Equation (\ref{#1})}
\begin{document}

\title[Half-Vortex Unbinding and Ising Transition in Constrained Superfluids]{Half-Vortex Unbinding and Ising Transition in Constrained Superfluids}

\author{Lars Bonnes$^{1,2}$ and Stefan Wessel$^{3}$}
\address{$^1$ Institut f\"ur Theoretische Physik III, Universit\"at Stuttgart, Pfaffenwaldring 57, D-70550 Stuttgart, Germany}
\address{$^2$ Institute for Theoretical Physics, University of Innsbruck, A-6020 Innsbruck, Austria}
\address{$^3$ Institute for Theoretical Solid State Physics, JARA-FIT,  and JARA-HPC, RWTH Aachen University, Otto-Blumenthal-Str. 26, D-52056 Aachen, Germany}
\ead{lars.bonnes@uibk.ac.at}

\date{\today}

\begin{abstract}
We analyze the thermodynamics of the atomic and (nematic) pair superfluids appearing in the  attractive
two-dimensional Bose-Hubbard model with a three-body hard-core constraint that has been derived as an effective model for cold atoms subject to strong three-body losses in optical lattices.
We show that the thermal disintegration of the pair superfluidity is governed by 
the proliferation of fractional half-vortices leading to a Berezinskii-Kosterlitz-Thousless transition with unusual jump in the helicity modulus.
In addition to the (conventional) Berezinskii-Kosterlitz-Thousless transition out of the atomic superfluid, 
we furthermore identify a direct thermal phase transition separating the pair and the atomic superfluid phases, and show that this transition is continuous with critical scaling exponents consistent with those of the two-dimensional Ising universality class.
We exhibit a direct connection between the partial loss of quasi long-range order at the Ising  transition between the two superfluids and the parity selection in the atomic winding number fluctuations that distinguish the atomic from the pair superfluid.
\end{abstract}

\section{Introduction}
\begin{figure}[t]
\begin{center}
\includegraphics[width=4.5cm]{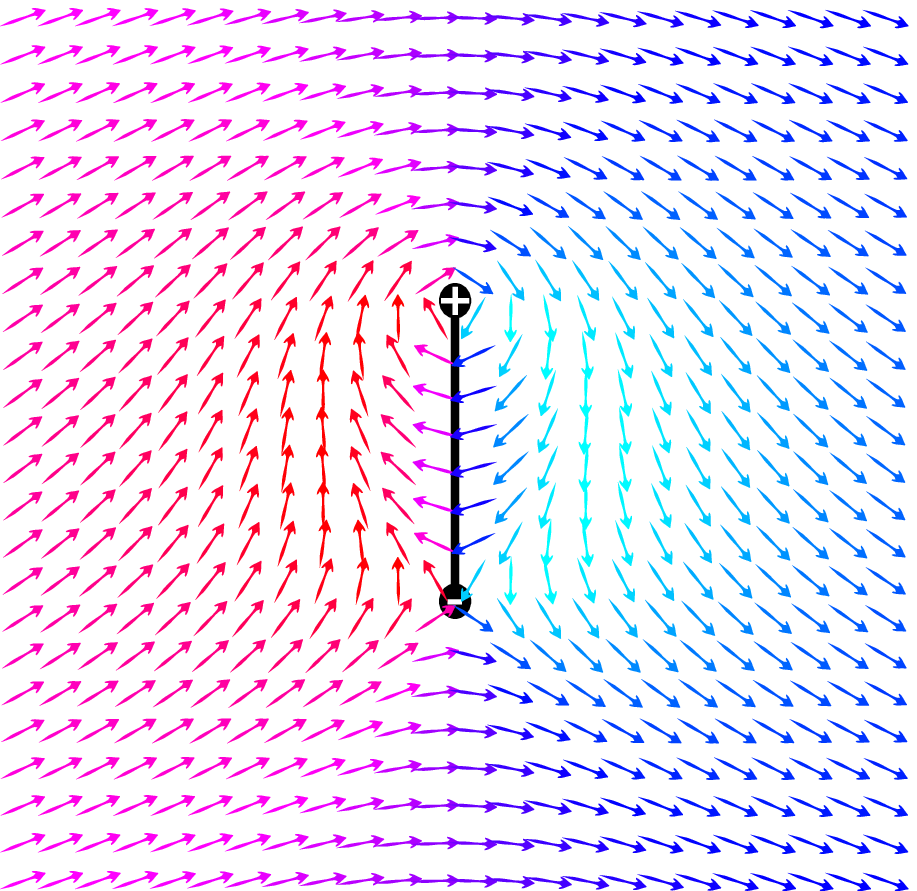}
\includegraphics[width=4.5cm]{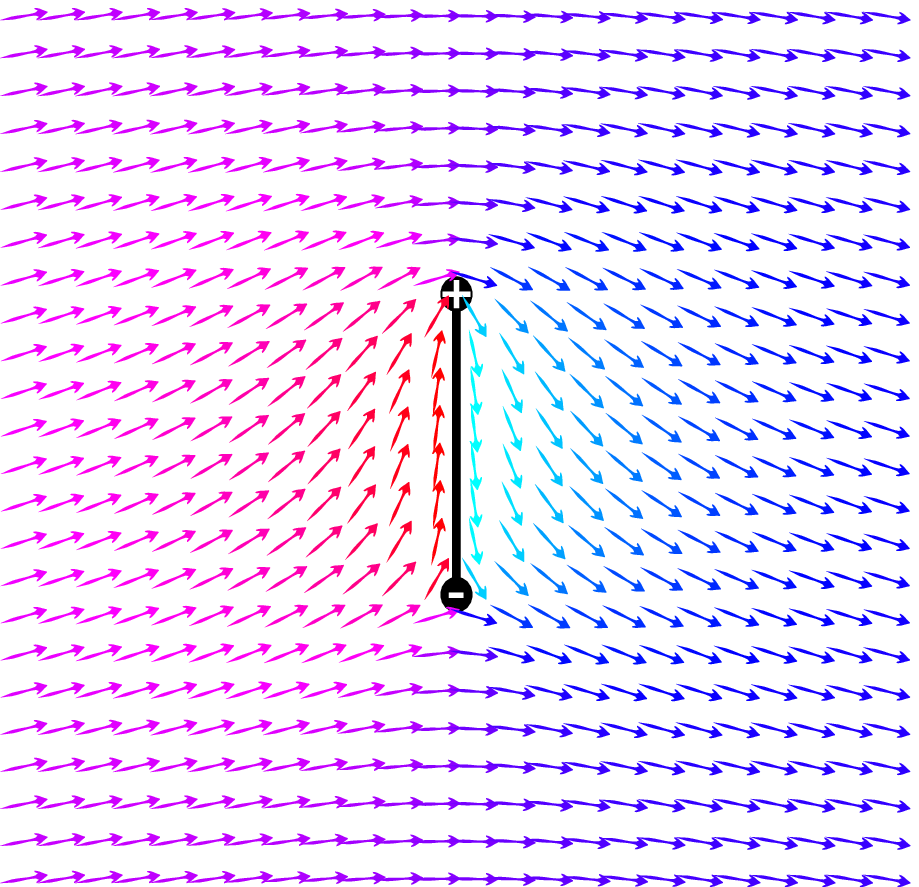}
\caption{The left (right) panel shows a (half-)vortex-antivortex pair in a classical $XY$ model, where the phase is represented by the direction of the shown arrows.
Whereas the arrows along the line separating the vortex-antivortex cores (denotes in the figure by + and -) are aligned, they form a domain wall (string) of anti-parallel  spins in the case of half-vortices.
}
\label{fig:vortex}
\end{center}
\end{figure}

Since the seminal work by Hohenberg~\cite{hohenberg67} and by Mermin and Wagner~\cite{mermin66}, it is well known, 
that true long-range order is destroyed at finite temperatures in two-dimensional short-range-interacting systems with a continuous symmetry, due to of the proliferation of order parameter fluctuations induced by  low energy excitation modes.  
However, quasi-long range order, characterized by an algebraic decay of correlations, can still persist at low temperatures, as is the case e.g. in the two-dimensional XY model, 
modelling the physics of e.g. coplanar magnets with $U(1)$ symmetry.
For the  finite temperature properties of the two-dimensional XY model, vortex and anti-vortex configurations, constituting topological defects in the field configuration, around which the XY order parameter acquires a phase difference of $2 \pi\nu_V$ with $\nu_V=\pm 1$ (cf. the left panel of \Fref{fig:vortex}) are essential, as their 
unbinding beyond the Berezinskii-Kosterlitz-Thousless (BKT) transition temperature~\cite{berezinskii71,kosterlitz73} destroys  the quasi-long-range order of the low-temperature phase.
An even richer phase structure can be found in generalizations of the XY model: Directly coupling to the vortex core energy upon introducing additional plaquette-couplings~\cite{swendsen82} or by introducing a nematic nearest-neighbor coupling term, half-vortices with fractional vorticity $\nu_V=\pm 1/2$ can be stabilzed~\cite{lee85,korshunov85,carpenter89}.
Half-vortices exhibit a phase winding of $\pi$ in the XY order parameter, and a half-vortex antivortex pair is  connected by a string with a finite tension formed by a domain wall, cf. the right panel of \Fref{fig:vortex}. For sufficiently weak string tension, a BKT transition of half-vortices can occur upon cooling the system down from high temperatures. Below this transition temperature, the domain-wall strings connecting the now confined half-vortex antivortex pairs form. They disappear at even lower temperatures, once the strings' tension dominates over their entropic contribution to the free energy. This second transition relates to a discrete $Z_2$ symmetry distinction between the low temperature phase and the (quasi-) nematic intermediate phase,  and  belongs to the universality class of the 
two-dimensional Ising model~\cite{lee85,korshunov85,carpenter89}.
Recently, classical XY models with generalized nematic couplings have attracted renewed interest~\cite{korshunov02,park08,poderoso11,shi11} not only because of  fractional vortex unbinding at high temperatures but also since the transitions between different quasi-long-range-ordered phases show intriguing critical properties~\cite{park08,poderoso11,shi11}.

\begin{figure}[t]
\begin{center}
\includegraphics[width=10cm]{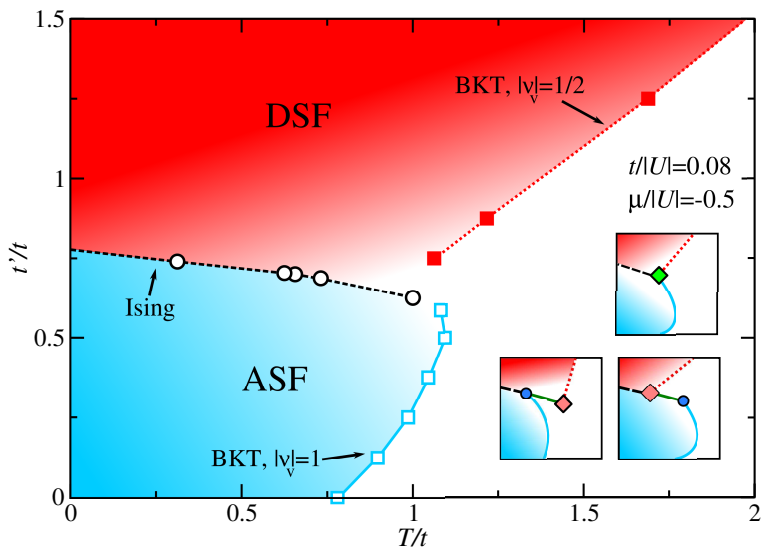}
\caption{Finite-temperature phase diagram of constrained bosons on the square lattice for $t/|U|=0.08$  and $\mu/|U|=-0.5$. 
The ASF-DSF transition belongs to the two-dimensional Ising universality class (dashed line).
The blue and red lines indicate the BKT transitions of the ASF and DSF, respectively and are distinguished by the unbinding of vortices ($\nu_V=1$) and half-vortices ($\nu_V=1/2$). 
The insets show different possible scenarios for the region where the two BKT lines and the Ising transition line meet.
}
\label{fig:phasediagram}
\end{center}
\end{figure}

Turning towards quantum systems, the conventional BKT-physics finds various realizations, e.g. by the phase fluctuations of a two-dimensional Bose gas near the superfluid-to-normal transition~\cite{landau49,feynman54} in Helium films and in two-dimensional quantum gases~\cite{bloch08}, where topological vortices have been directly imaged~\cite{yarmchuck79,madison00,madison00a,shaeer01,hadzibabic06}. Furthermore, 
half-vortex excitations have been predicted~\cite{rubo07,flayac10,solano10,assmann10} and were indeed observed~\cite{lagoudakis09} in exciton-polariton condensates. 
Spinor condensates in the presence of spin-orbit coupling are also expected to give rise to the formation of half-vortex states~\cite{ramachandhran12}.
Recently, attention has focused on gases of attractive bosons, 
where the above scenario of half-vortex topological defects and strings is realized due to the emergence of two different kinds of superfluid phases: besides a conventional atomic superfluid, a molecular superfluid of boson pairs (dimers) can form, which provides half-vortex topological defects~\cite{romans04,radzihovsky04}. 
To prevent the attractive Bose gas from collapse, a scenario has been proposed recently, wherein the Bose gas is constrained to an optical lattice and furthermore subject to strong three-body losses that project out triple occupancy on each lattice site~\cite{daley09}. 
The dynamical suppression of triply occupied sites has been demonstrated eventually in Cesium condensates~\cite{mark12}.
The effective model that describes this constrained lattice boson system is the 
Bose-Hubbard model with a onsite three-body constraint, $(b_i^\dagger)^3=0$, given by the Hamiltonian
\begin{equation}
 H = - t \sum_{\nn{ij}} ( b_i^\dagger b_j + \hc ) 
- t' \sum_{\nn{ij}} [ (b_i^\dagger)^2 (b_j)^2 + \hc ] 
+ \frac{U}{2} \sum_i n_i (n_i-1)
- \mu \sum_i n_i,
\label{eq:hamiltonian}
\end{equation}
where $b_i$ ($b_i^\dagger$) are bosonic annihilation (creation) operators and $n_i=b_i^\dagger b_i$.
Here, $t$ and $t'$ denote hopping terms of atoms and dimers between nearest-neighbors $\nn{ij}$, $U<0$ is the attractive on-site interaction and the filling is controlled by a chemical potential $\mu$.
Furthermore, the bosonic operators have to fulfill the constraint $(b_i^\dagger)^3=0$, restricting the local occupation to two at most.
For $t'=0$, this model has been subject to extended analytical~\cite{daley09,diehl10,diehl10a,diehl10b,lee10} and numerical~\cite{bonnes11b,ng11,chen11} studies.
We are interested here in particular in the case of a two-dimensional square lattice geometry. 
Besides the atomic superfluid phase (ASF) with an atomic quasi-condensate (showing an algebraic decay of the atomic Green's function $\nn{b^\dagger_i b_j}$), a pair superfluid  phase proliferates in the strong coupling regime ($t \ll |U|$) where the atomic correlations decay exponentially, but the  boson pair correlation function $\nn{(b^\dagger_i)^2 (b_j)^2}$ still decay algebraically. This distinct superfluid phase is also called the  dimer superfluid phase (DSF).
Recent quantum Monte-Carlo (QMC) calculations exhibited, that the finite temperature superfluid transition to the high-temperature phase is governed by the unbinding of half-vortices in the case of the DSF~\cite{bonnes11b,ng11} such that this system provides an interesting candidate for the investigation of such topological excitations in state-of-the-art experiments.
The DSF phase is distinct from the ASF phase by a residual $Z_2$ invariance, similar to the nematic phase of the aforementioned classical models~\cite{romans04,radzihovsky04}. Here, we explore in particular the possibility of a direct thermal phase transition between the ASF and DSF region, and study its critical  properties.

For this purpose, we analyze the Hamiltonian of \Eqref{eq:hamiltonian} for which the ASF-DSF transition is driven explicitly upon varying the pair hopping amplitude $t'$. Such correlated hopping terms have been derived also for spin-1 bosons within an effective three-body constrained Hubbard model description~\cite{mazza10}. Using large scale QMC simulations, we focus here on exploring the thermal phase transitions in this model between the ASF, DSF and normal fluid (NF) phases. 
The transition from the ASF to the DSF phase can also be driven upon increasing the interaction strength $|U|$ instead of  $t'$, and one would expect the universal properties of the considered phase transitions to not depend on the choice of the driving parameter. 
However, since the QMC algorithm performs very efficiently in the presence of a finite pair hopping term $t'$, we can access  
considerably larger system sizes than for the $|U|$-driven transitions at $t'=0$~\cite{bonnes11b}. 

After presenting methodological aspects in the following section, we reveal in Section 3 the presence of a continuous direct ASF-DSF thermal phase transition, and provide evidence, that this transition belongs to the two-dimensional Ising universality class, related to the residual $Z_2$ symmetry distinction between the ASF and DSF phase.  In particular, we identify relevant critical quantities and the related critical exponents, and show that 
the helicity modulus from the odd-parity subsector, with a scaling dimension of zero, can be used to assess the critical point rather accurately.
In Section 4, we then
provide 
a detailed analysis of the  ASF-NF and the DSF-NF thermal phase transitions. 
This allows us to clearly contrast the usual vs. unusual nature of the BKT transitions 
for the ASF and DSF phase, respectively from simulations on lattices with a linear size extending up to $L=200$. 
The finite-temperature phase diagram, shown in \Fref{fig:phasediagram}, summarizes the results of our analysis and will be further discussed in the following sections.

\section{Method}
The thermodynamic properties of the model presented in \Eqref{eq:hamiltonian} are here studied using large-scale QMC simulations;
we employ a generalized directed loop algorithm in the stochastic series representation~\cite{sandvik99b,syljuasen02,alet05} on a square lattice geometry with linear system size $L$ and $N=L^2$ lattice sites. In order to account for the dimer hopping term in \Eqref{eq:hamiltonian}, we use two types of directed loops, where the worm can either carry a single boson or a boson pair (dimer) creation/annihilation operator.
Typical system sizes simulated in this work range from $L=10$ to $200$ and allow for a precise finite-size analysis of the data, as discussed below. The simulations are performed at finite temperatures $T$, and with periodic boundary conditions taken along both spatial directions.

To access the superfluid response of the system, 
we calculate the helicity modulus
\begin{equation}
 Y = \frac{T \nn{\mathbf{W}^2}}{2 |U|},
\end{equation}
by measuring the winding number fluctuations $\mathbf{W}=(W_x, W_y)$ along the $x$- and $y$-directions. This quantity relates directly to the superfluid stiffness, and thus allows to detect the onset of superfluidity at the BKT transition~\cite{pollock87,harada97,bernardet02}. 
Furthermore, we define even and odd helicities, $Y_\mathrm{even/odd}$, that  take into account only winding numbers from the corresponding parity sector, i.e. 
for $Y_\mathrm{even}$ ($Y_\mathrm{odd}$) both $W_x$ and $W_y$ must be even (odd).
These parity-restricted helicities allow for a discrimination of the ASF and DSF phases~\cite{schmid06,bonnes11b,ng11}, as discussed below. 
Note that a factor $1/|U|$ is included to make $Y$   dimensionless in the above formula.
The quasi-long-range order within the ASF and DSF phases are characterized by structure factors which correspond to 
the atomic and dimer condensate densities, $C_1$ and $C_2$, respectively,  which are measured via the atomic and dimer equal-time Green's functions 
\begin{equation}
G_1(i,j)=\nn{b_i^\dagger b_j}, \quad G_2(i,j)=\nn{ (b_i^\dagger)^2 (b_j)^2}
\end{equation}
in the standard way during the off-diagonal directed-loop update~\cite{alet05,dorneich01} and are defined as
\begin{equation}
C_a = \frac{1}{N^2} \sum_{ij} G_a(i,j),
\end{equation}
for $a=1,2$. 
Previous numerical simulations exhibited a sampling problem  of the dimer Green's function in the limit  $t'=0$, originating from the appearance of a fat-tailed distribution in the histogram of the observable for $G_2$ inside the DSF phase~\cite{bonnes11b}.
The following simulations, however, are performed in a regime where the system is well within the ASF phase for $t'=0$, such that these sampling problems could be avoided.

\section{Ising Transition}

\begin{figure}[t]
\begin{center}
\includegraphics[width=10cm]{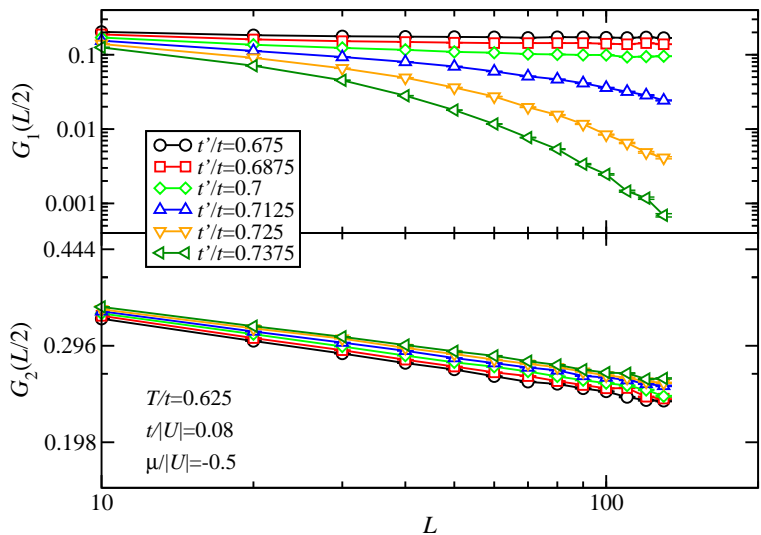}
\caption{Atomic ($G_1$) and dimer ($G_2$) equal-time Greens's function at the largest distance $L/2$
for different values of the dimer hopping strength $t'$, at fixed $T/t=0.625$.}
\label{fig:g1g2}
\end{center}
\end{figure}

The phase diagram for the model considered in \Eqref{eq:hamiltonian}, shown in \Fref{fig:phasediagram}, was obtained for fixed values of $t/|U|=0.08$ and $\mu/|U|=-0.5$.
These values were chosen such that at low temperatures the system is still in the ASF phase for $t'=0$ (the ASF-DSF quantum phase transition for $t'=0$ is located at $(t/|U|)_c \approx 0.055$~\cite{bonnes11b}), and thus exhibits quasi-long-range-order in both the atomic and dimer channel as well as both even and odd winding numbers.

Upon increasing $t'/t$ at low temperatures, the system eventually enters the DSF phase, due to the enhanced dimer hopping strength $t'$. Indeed, beyond a critical value of $t'/t\approx 0.7$, 
the atomic Green's function ceases to exhibit algebraic scaling, while the pair Green's function still remains critical. This can be seen in \Fref{fig:g1g2}, which shows both $G_1$ and $G_2$ at the largest distance $L/2$ as a function of the linear system size $L$ for various values of $t'/t$ around 0.7 at a temperature of $T/t=0.625$. While the pair  Green's function  $G_2$ remains critical throughout this regime, the atomic  Green's function  $G_1$ exhibits a 
clear suppression at large length scales for  $t'/t$ beyond 0.7, indicating the breakdown of its characteristic algebraic scaling in the ASF phase. 
The ASF-DSF transition has a slight temperature dependence (see \Fref{fig:phasediagram}) such that the system can be driven  from one superfluid regime to the other one via controlling the temperature.
As discussed above, one expects this finite temperature direct ASF-DSF transition to be within the two-dimensional Ising universality class.

\begin{figure}[t]
\begin{center}
\includegraphics[width=10cm]{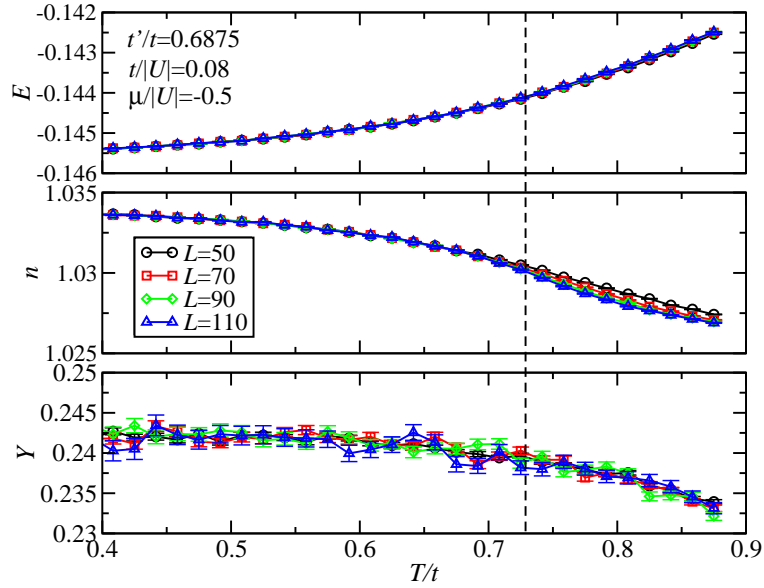}
\caption{Energy $E$, filling $n$ and helicity $Y$ as functions of the temperature $T$ for $t'/t=0.6875$, providing no indication for first-order behavior at the ASF-DSF transition at $(T/t)_c \approx 0.73$, indicated by the dashed line.
}
\label{fig:e_and_n}
\end{center}
\end{figure}

To assess the nature of the finite-$T$ ASF-DSF transition, we first consider the behavior of various  thermodynamic quantities across the transition line. 
In particular, \Fref{fig:e_and_n} shows the energy $E$, filling  $n$ and total helicity $Y$ for $t'/t=0.6875$, all of which are seen to vary smoothly across the thermal transition located at $(T/t)_c\approx0.73$.
Beyond this temperature, i.e. within the DSF region, the filling $n$ shows weak but visible finite size effects, indicative of a different state as compared to the low-$T$ phase. 
Based on the smooth behavior of all these quantities across the transition, we 
do not obtain any evidence for a first-order transition. We thus continue to analyze the direct ASF-DSF transition in terms of 
its critical properties.

\begin{figure}[t]
\begin{center}
\includegraphics[width=10cm]{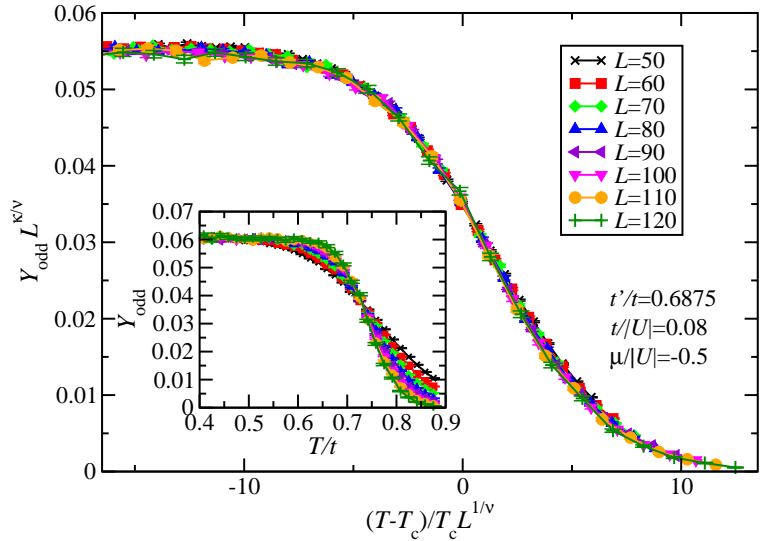}
\caption{Data collapse for the odd helicity $Y_\mathrm{odd}$ for $t'/t=0.6875$ using $\kappa=0$ and $\nu=1$. The
inset shows $Y_\mathrm{odd}$ for different system sizes across the critical point at $(T/t)_c \approx 0.73$.
}
\label{fig:scalingYodd}
\end{center}
\end{figure}

For this purpose, we first focus on the odd-parity helicity $Y_\mathrm{odd}$, which exhibits distinct behavior in the  ASF vs. DSF phase:
The inset of \Fref{fig:scalingYodd} shows  $Y_\mathrm{odd}$ as a function of $T$ for $t'/t=0.6875$. This quantity
indeed exhibits a pronounced change in its finite-size scaling behavior at about $T/t \approx 0.73$: at lower $T$, it increases with system size, approaching an essentially $T$-independent finite value in the thermodynamic limit, while at larger $T$, it scales to zero in the thermodynamic limit. 
Furthermore, $Y_\mathrm{odd}$ shows a crossing point of the finite-size data at criticality.
This is reminiscent of the general scaling behavior of  helicities (or superfluid densities) at second-order thermal phase transitions, i.e. $Y_\mathrm{odd} \propto L^{2-d}$.
In two dimensions ($d=2$) $Y_\mathrm{odd}$ thus has scaling dimension zero, similar to the Binder ratio~\cite{binder81,binder02}, and can hence be used for a precise estimation of the critical point.
This critical scaling behavior of $Y_\mathrm{odd}$ is confirmed by applying a standard scaling ansatz
\begin{equation}
Y_\mathrm{odd}  = L^{-\kappa /\nu} g\left[ \frac{T-T_c}{T_c} L^{1/\nu}\right].
\end{equation}
Here, $\kappa$ and $\nu$ denote critical exponents and $g$ is a scaling function.
In order to extract the critical exponents as well as $T_c$, we fit the finite-size data for $Y_\mathrm{odd}$ to the above scaling ansatz, approximating the scaling function $g$ by a fourth-order polynomial.
The error bars for the fitting parameters are estimated using a standard bootstrap analysis~\cite{shao95}.
From this analysis, we obtain the values $(T/|U|)_c=0.726 \pm 0.006$,  $\nu=1.08 \pm 0.09$ and $\kappa=-0.01 \pm 0.04$, consistent with the two-dimensional Ising critical exponent $\nu=1$~\cite{pelissetto02} and with $\kappa=0$. Indeed, a robust data collapse of $Y_\mathrm{odd}$ is observed in the main panel of \Fref{fig:scalingYodd} using the Ising  exponent  $\nu=1$ and $\kappa=0$.

\begin{figure}[t]
\begin{center}
\includegraphics[width=10cm]{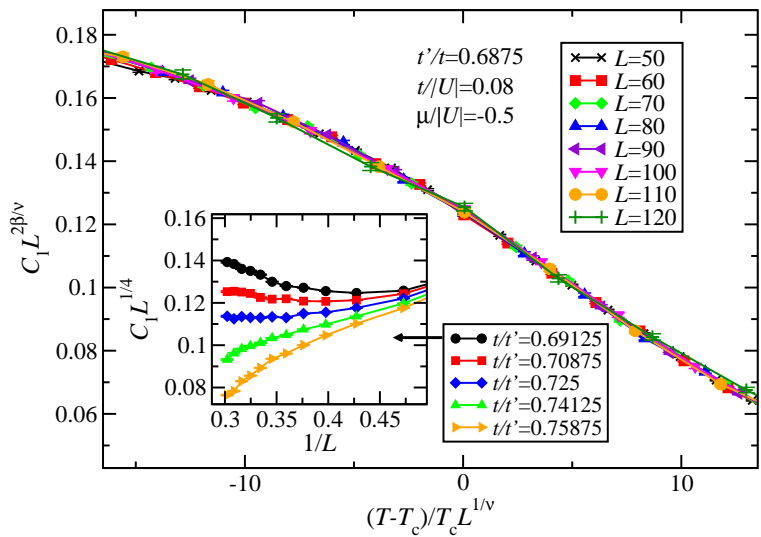}
\caption{Data collapse for the condensate density $C_1$ for $t'/t=0.6875$ using standard Ising critical exponents ($\beta=1/8$ and $\nu=1$). 
Inset: Rescaled condensate density $C_1$ in the vicinity of the critical point exhibiting its algebraic decay $C_1 \propto L^{-1/4}$.
}
\label{fig:scalingc1}
\end{center}
\end{figure}

Further confirmation of  Ising criticality  at the finite-$T$ ASF-DSF transition is obtained from analyzing the atomic condensate density $C_1$.
While in two dimensions at finite $T$, the expectation value of the atomic condensate density vanishes in both phases, it still exhibits  characteristic algebraic scaling within the transition region. 
In particular, the inset of \Fref{fig:scalingc1} shows the finite-size behavior of $C_1$ multiplied by $L^{1/4}$, which makes evident  characteristic scaling $C_1\propto L^{-1/4}$ in the vicinity of the critical temperature.
Using a standard scaling ansatz 
\begin{equation}
C_1  = L^{-2\beta/\nu} h\left[ \frac{T-T_c}{T_c} L^{1/\nu}\right],
\end{equation}
with a scaling function $h$, the above scaling of $C_1\propto L^{-1/4}$ at $T=T_c$ is consistent with the critical exponents $\beta$ and $\nu$ taking on their two-dimensional Ising values $\beta=1/8$ and $\nu=1$~ \cite{pelissetto02}.
Performing an unbiased fit of the finite-size data to the above scaling ansatz for $C_1$, we obtain  $(T/|U|)_c=0.728 \pm 0.009$, $\beta=0.125 \pm 0.005$ , and $\nu=0.95 \pm 0.07$. These values are indeed consistent with the Ising model values and the value of $\nu$ is also consistent with the result from the scaling analysis of $Y_\mathrm{odd}$. The main panel of \Fref{fig:scalingc1} show the data collapse of $C_1$, using exact Ising exponents.
The critical exponents found for this specific case of $t'/t=0.6875$ are consistent also with values obtained along other cuts through the ASF-DSF phase transition line.
We thus conclude, that at finite temperatures a direct ASF-DSF transition is present, that belongs to the two-dimensional Ising universality class. Furthermore, the  odd-parity helicity $Y_\mathrm{odd}$, being a dimensionless quantity, provides a direct connection between the emergence of the $Z_2$ symmetry above the ASF-DSF transition line and the loss of bare atomic contributions to the (odd) single-particle winding beyond the ASF region.

\section{BKT Transitions}

\begin{figure}[]
\begin{center}
\includegraphics[width=10cm]{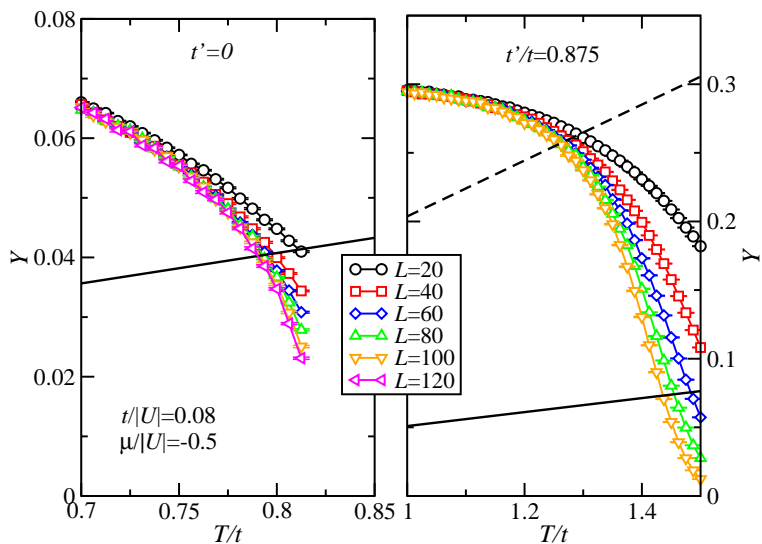}
\caption{Helicity $Y$ as a function of temperature $T/t$ for different system sizes for $t'=0$ (left panel) and $t'/t=0.875$ (right panel) at $t/|U|=0.08$ and $\mu\|U|=-0.5$. The solid (dashed) line relates to the usual (unusual) universal helicity jump $2T/(\pi|U|)$, ($8T/(\pi|U|)$). 
}
\label{fig:sfds}
\end{center}
\end{figure}

We next consider the phase transitions from the ASF and DSF phases into the NF high temperature regime. For both 
the ASF and DSF phases, the system  undergoes a BKT transition at a finite temperature $\Tbkt$.
A standard way of estimating $\Tbkt$ for a conventional superfluid stems from the universal jump of the helicity modulus~\cite{nelson77} and knowledge of the finite-size scaling behavior of the helicity at the BKT point, as derived from the BKT renormalization group equations~\cite{weber88},
\begin{equation}
 Y(T, L) = \frac{2 T}{\pi |U|}A 
 \left( 1 + \frac{1}{2 \log[L/L_0]} \right).
 \label{eq:YTL}
\end{equation}
Here, $L_0$ is of order unity and $A=1/\nu_V^2$ is determined by the elementary vorticity $\nu_V$ at $T=\Tbkt$.
Previous studies of the thermodynamic properties at $t'=0$ showed that the ASF undergoes a conventional BKT transition with an unbinding of integer vortices ($\nu_V=1$), for which $A=1$, whereas the DSF shows the unbinding of half-integer vortices ($\nu_V=1/2$) and the universal jump of the helicity modulus is thus increased by a factor of four ($A=4$)~\cite{bonnes11b,ng11}.

This can be seen also for the present case. In \Fref{fig:sfds}, the helicity modulus $Y$ is shown  for various system sizes for two different values of $t'/t$, corresponding to the ASF ($t'=0$) and DSF ($t'/t=0.875$) phase, respectively. While for the ASF case, the data is consistent with the usual helicity jump of $2T/(\pi|U|)$, the data for the DSF case exhibits a behavior that is consistent only with the larger helicity jump  of $8T/(\pi|U|)$ at the BKT transition.

In order to confirm this behavior in more detail, we now employ the above scaling formula:
Since this scaling form of the helicity modulus is valid only right at the BKT transition temperature, the identification of the best fit of the finite-size data to \Eqref{eq:YTL} indeed allows for a precise determination of the actual  helicity jump, from which the vorticity $\nu_V$ can be inferred, as has been demonstrated previously, cf. e.g. Refs.~\cite{korshunov02,park08}.
As a measure of the goodness-of-fit, we use $\chi^2/\mathrm{DOF}$~\cite{press92}, where $\mathrm{DOF}$ denotes the number of degrees of freedom of the fit and 
\begin{equation}
 \chi^2 = \sum_{L_i} \left( \frac{ Y(T,L_i) - Y^\mathrm{fit}(T,L_i)}{\sigma_i} \right)^2
\end{equation}
is performed by summing over a range of finite system sizes $L_i$.
Here, $\sigma_i$ is the Monte Carlo error of $Y(T, L_i)$, and $Y^\mathrm{fit}$ denotes the fitting function in \Eqref{eq:YTL}. 
\begin{figure}[]
\begin{center}
\includegraphics[width=10cm]{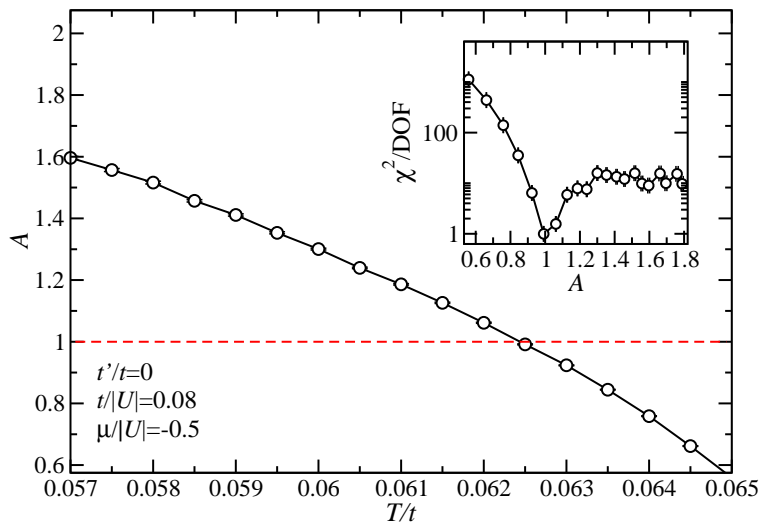}
\caption{$A$ versus $T/t$ for $t'=0$ and $t/|U|=0.08$ and $\mu\|U|=-0.5$. The inset shows  $\chi^2/\mathrm{DOF}$ vs. $A$ for the fit of the finite-size data using system sizes $L=10,\,20,\,30,\,..,\,200$ to \Eqref{eq:YTL}.
The goodness of the fit shows a clear minimum at $A=1$, indicating the unbinding of integer vortices.
}
\label{fig:tprime0}
\end{center}
\end{figure}
We first illustrate this procedure for the transition at $t'=0$ from the ASF to the NF. 
\Fref{fig:tprime0} shows $A$ and $\chi^2/\mathrm{DOF}$ for $t'=0$, $t/|U|=0.08$ and $\mu/|U|=-0.5$.
The pronounced minimum in $\chi^2/\mathrm{DOF}$ in the best fit for  $A=1$ reproduces the standard value $\nu_V=1$ for a two-dimensional Bose-Hubbard model out of the ASF phase.
In this case, we find that the position of the minimum in $\chi^2/\mathrm{DOF}$  does not depend on the range of the system sizes that is included in the fitting procedure.

\begin{figure}[]
\begin{center}
\includegraphics[width=10cm]{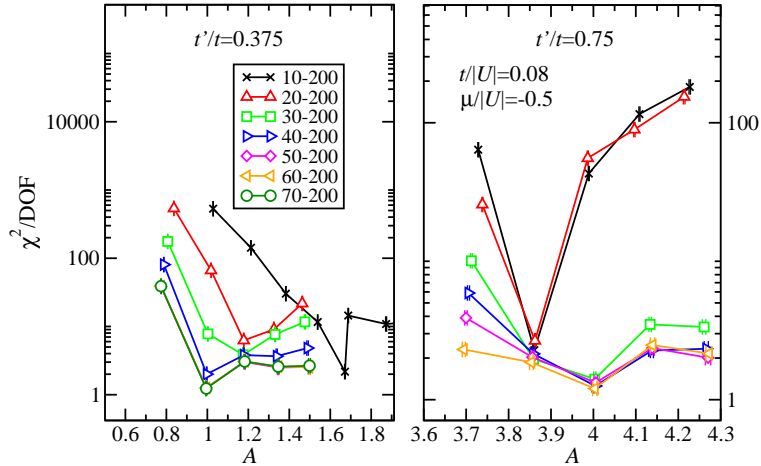}
\caption{Goodness of the fit ($\chi^2/\mathrm{DOF}$) versus $A$ using different fitting ranges for $t'/t=0.375$ (left) and $t'/t=0.75$ (right) at $t/|U|=0.08$ and $\mu\|U|=-0.5$.
The minimum of the goodness, $\chi^2/\mathrm{DOF}$, is shifted towards $A=1$ or $4$ as smaller system sizes are omitted from the fitting procedure.
}
\label{fig:tprime0.6}
\end{center}
\end{figure}

For finite values of $t'/t$, however, we observe a mild dependence of the best fit value of $A$ (i.e. the position of the minimum in $\chi^2/\mathrm{DOF}$) on the considered range of system sizes.
We thus  need to carefully assess the dependence of the fitting result, in particular the best-fit value of $A$, on the considered range of system sizes, since \Eqref{eq:YTL} only holds within the scaling regime, i.e. sufficiently large values of $L$ are available for the fitting procedure. 
In order to extract for a given ratio $t'/t$ the best fit value of $A=1/\nu_V^2$ from the minimum in $\chi^2/\mathrm{DOF}$, we thus monitored the
dependence of the fitting procedure on the employed range of system sizes. 
This is illustrated for two different cases, $t'/t=0.375$ (ASF) and $0.75$ (DSF), in \Fref{fig:tprime0.6}. One  observes that (i) the minimum of $\chi^2/\mathrm{DOF}$ shifts towards $A=1$ (ASF) or $4$ (DSF) as more and more of the smaller system sizes are omitted from the fit, and (ii) that the best-fit value of $A$ converges eventually.
Thus, the ASF and DSF phases are characterized by $\nu_V=1$ and $\nu_V=1/2$, respectively and $\nu_V$ does not show  crossover behavior as might  have been inferred from a naive scaling analysis without monitoring the system-size-range convergence. 
Performing the above analysis for different values of $t'/t$, we eventually arrive at 
the phase boundaries for the ASF and DSF phases shown in \Fref{fig:phasediagram}.
It is noteworthy that $\Tbkt$ is suppressed in the vicinity of the ASF-DSF transition and gives rise to a superfluid re-entrance phenomenon where the system, upon increasing $t'/t$ at fixed $T\approx t$, goes from the ASF to the NF phase and eventually enters the DSF upon further increasing $t'$.

\section{Conclusion}
In conclusion, we analyzed the finite-temperature attractive Bose-Hubbard model with a three-body hard-core constraint on the square lattice using large-scale QMC calculations.
The transition to a (quasi-) nematic DSF is driven here by explicitly adding a nearest-neighbor pair hopping term, which allows for efficient simulations using a two-worm algorithm.
The pair-hopping driven transition between the two quasi-ordered superfluids is found to be in the Ising universality class, accessible from the behavior of the dimensionless 
odd-parity helicity $Y_\mathrm{odd}$, having a scaling dimension of zero in two dimensions.
For the case $t'=0$, where the DSF proliferates in the strong-coupling regime, $|U|\gtrsim 20 t$, Ng and Yang obtained evidence for a strong first-order direct ASF-DSF transition at finite temperatures~\cite{ng11}. 
In addition to the fact that their model includes an additional nearest-neighbor repulsion, it is  possible that the Ising transition will eventually be driven first-order as one approaches the strong-coupling regime in which the physics will be controlled by an effective Feshbach model~\cite{romans04,radzihovsky04,radzihovsky08} with a sizeable Feshbach detuning $\sim |U|$, as derived in Ref. \cite{diehl10b}.

An open  question concerns the topology of the phase diagram (see \Fref{fig:phasediagram}) in the region, where the two BKT transition lines and the Ising transition line meet.
In the insets in \Fref{fig:phasediagram}, we have drawn different possible scenarios, wherein the three lines either meet at a multicritcal point or where  another (possibly Ising or first-order) transition line persists between one of the superfluid regions and the high temperature phase.
Identifying the precise nature of the phase structure in this region is beyond the scope of this work and we leave this interesting issue  for future research.

\section*{Acknowledgements}
We acknowledge discussions with S. Diehl, H. G. Evertz, A. L\"auchli, K.-K. Ng, L. Radzihovsky, and K. P. Schmidt.
The numerical simulations were performed on the high performance computers at HLRS Stuttgart and NIC J\"ulich from whom we acknowledge the allocation of CPU time.
This work was supported by the Studienstiftung des Deutschen Volkes and the DFG within SFB/TRR 21.

\section*{References}
\bibliographystyle{unsrt}

\end{document}